\documentclass[prl,twocolumn,preprintnumbers,amsmath,amssymb,showkeys,showpacs]{revtex4}
\usepackage[pdftex]{graphicx}
\usepackage{dcolumn}
\usepackage{bm}
\begin{document}
\title{Ultrahigh electron mobility in suspended graphene}
\author{K. I. Bolotin$^a$}
\author{K. J. Sikes$^b$}
\author{Z. Jiang$^{a,d}$}
\author{M. Klima$^{c}$}
\author{G. Fudenberg$^a$}
\author{J. Hone$^c$}
\author{P. Kim$^a$}
\author{H. L. Stormer$^{a,b,e,*}$}

\affiliation{Departments of $^a$Physics, $^b$Applied Physics, $^c$Mechanical Engineering, Columbia University, New York, NY 10027, USA}
\affiliation{$^d$National High Magnetic Field Laboratory, Tallahassee, FL 32310, USA}
\affiliation{$^e$Bell Labs, Alcatel-Lucent Technologies, Murray Hill, NJ 07974, USA}
\date{\today}

\begin{abstract}
We have achieved mobilities in excess of 200,000 cm$^2$V$^{-1}$s$^{-1}$ at electron densities of $\sim$2$\times$10$^{11}$ cm$^{-2}$ by suspending single layer graphene. Suspension $\sim$150 nm above a Si/SiO$_2$ gate electrode and electrical contacts to the graphene was achieved by a combination of electron beam lithography and etching. The specimens were cleaned in situ by employing current-induced heating, directly resulting in a significant improvement of electrical transport. Concomitant with large mobility enhancement, the widths of the characteristic Dirac peaks are reduced by a factor of 10 compared to traditional, non-suspended devices. This advance should allow for accessing the intrinsic transport properties of graphene.
\end{abstract}

\maketitle
Graphene, the latest addition to the family of two-dimensional (2D) materials, is distinguished from its cousins by its unusual band structure, rendering the quasiparticles in it formally identical to massless, chiral fermions. The experimental realization of graphene thus presents tantalizing opportunities to study phenomena ranging from the topological phase resulting in exotic quantum Hall states \cite{novoselov,yuanbo} to the famous Klein paradox -- the anomalous tunneling of relativistic particles \cite{rise_graphene}. However, despite tremendous interest and concerted experimental efforts [1-23], the presence of strong impurity scattering -- which limits the electron mean free path to less than a micron -- has been a major barrier to progress. At the same time, there is strong evidence that graphene is a nearly perfect crystal free of the structural defects \cite{elena,ishigami} that characterize most conductors. As a result, it has been put forth that the scattering of charge carriers stems from extrinsic sources \cite{nomura,dassarma,electrostatic,geim_intr}.

Although the exact nature of the scattering that limits the mobility of graphene devices remains unclear, evidence has mounted that interactions with the underlying substrate are largely responsible. Surface charge traps \cite{chencharged, dassarma, nomura,electrostatic}, interfacial phonons \cite{chen_limits}, substrate stabilized ripples \cite{suspend_geim, ishigami,geim_intr}, and fabrication residues on or under the graphene sheet may all contribute. Consequently, improving substrate quality or eliminating the substrate altogether by suspending graphene over a trench seems a promising strategy towards higher quality samples.  While devices suspended over the substrate were achieved in the past \cite{suspend_geim,bunch}, they lacked multiple electrical contacts thus precluding transport measurements.

In this Letter we report the fabrication of electrically contacted suspended graphene and achieve a tenfold improvement in mobility as compared to the best values reported in the literature for traditional devices fabricated on a substrate. Besides opening new avenues for studying the intrinsic physics of Dirac fermions, this improvement demonstrates the dominant role played by extrinsic scattering in limiting the transport properties of unsuspended graphene samples.

The fabrication of a suspended graphene device starts with optically locating a single-layer mechanically exfoliated graphene flake on top of a silicon substrate covered with 300 nm of SiO$_2$. Single-layer graphene flakes are identified based on their contrast \cite{geim_contrast}, and later confirmed via measurements of the half-integer quantum Hall effect \cite{yuanbo,novoselov}. We avoid patterning the flakes using oxygen plasma etching \cite{melinda, novoselov}, as it may introduce additional defects in the bulk and dangling bonds at the edges of graphene.  Instead, we choose natural flakes of approximately rectangular shape suitable for fabrication into Hall bars. Electron beam lithography is employed to pattern the contacts to the flake. The contact material (3 nm Cr followed by 100 nm of Au) is deposited by thermal evaporation followed by a liftoff in warm acetone. The large size and thickness of the electrodes enhances the mechanical rigidity of the device. Suspension of the graphene flake is achieved by dipping the entire device into 1:6 buffered oxide etch (BOE) for 90 seconds, which uniformly removes approximately 150 nm of SiO$_2$ across the substrate, {\it including the area below the flake} (SiO$_2$ masked by the gold electrodes remains unetched). Uniform etching of the substrate directly below the flake is crucial for our process as it allows the fabrication of large-area suspended graphene, while maintaining the parallel plate capacitor geometry for our device. To our knowledge, this unexpected etching anisotropy in the presence of graphene was not reported before; it is, however, consistent with the rapid propagation of BOE along the SiO$_2$/graphene interface \cite{me_unpublished}. Finally, the device is transferred from BOE to ethanol and dried in a critical-point-drying step to avoid the surface-tension-induced collapse of the suspended graphene sheet.

\begin{figure}
\includegraphics[width=80mm]{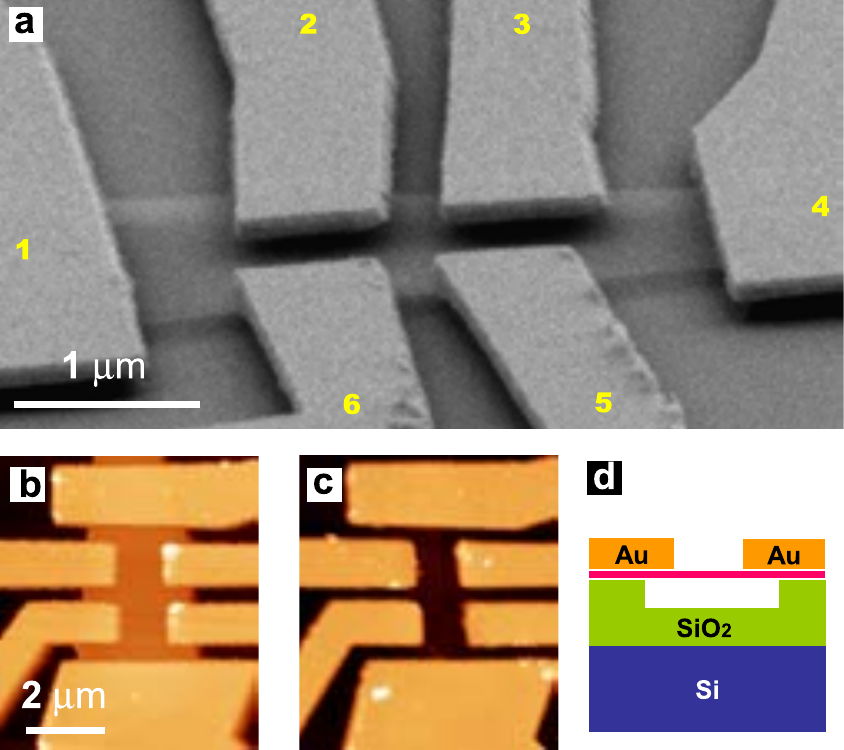}
\caption{(a) SEM image of a typical suspended six-probe graphene device taken at $15^{\circ}$ with respect to the sample plane. (b) AFM image of the suspended device \#1 before the measurements. (c) AFM image of the device \#1 after the measurements with graphene removed by a short oxygen plasma etch (same z scale). (d) Device schematic, side-view. Degenerately doped silicon gate (blue), partly etched SiO$_2$ (green), suspended single-layer graphene (pink) and Au/Cr electrodes (orange).}
\label{fig1}
\end{figure}

Figure 1a shows a scanning electron microscope (SEM) image of a finished device taken at $15^{\circ}$ angle with respect to the sample plane. The graphene is apparent as a thin sheet suspended above the surface of the remaining SiO$_2$. The sheet is supported by six gold electrodes attached to SiO$_2$, which have been slightly undercut during the BOE etching step (see Fig. 1d). Atomic force microscopy (AFM) (Figs. 1b,c) demonstrates convincingly the integrity of the graphene sheet, its suspension above the oxide and the flatness of the substrate below it. Fig. 1b clearly indicates a flat graphene surface $\sim$150 nm above the surface of SiO$_2$. The single layer of carbon atoms, which makes up graphene, is remarkably robust and is not damaged by repeated AFM imaging. Fig. 1c show the same device after completion of the electrical measurement and after removal of the suspended graphene via an oxygen plasma etch \cite{o2etch}. It reveals the previously hidden SiO$_2$ substrate below the graphene.  The height variation of the substrate is less than 20 nm, with a slight bowing towards the center of the device. We thus conclude that our fabrication process results in graphene devices suspended $\sim$150 nm above SiO$_2$ substrate (Fig. 1d).

Electrical measurements on suspended graphene devices are performed in a sample-in-vacuum cryostat with a pressure of less than $5\times 10^{-5}$ mtorr. A total of one four-probe and two six-probe devices were measured. Before cooling the cryostat to its base temperature of $\sim$5 K the devices are thermally annealed in situ to 400 K, as this has been shown to reduce spurious doping in unsuspended samples \cite{ishigami,schendin}. Four-probe measurements are performed using standard low-frequency lock-in techniques with the excitation current less than $I=100$~nA. A typical measurement consists of sending the current between electrodes labeled 1 and 4 in Fig. 1a and recording the voltages $V_{xx}$ ( $V_{xy}$ ) between electrodes 2 and 3 ( 2 and 6 ) respectively. The resistance is calculated as $R_{xx} = V_{xx}/I$ and the Hall resistance as $R_{xy}= V_{xy}/I$. To convert resistance to resistivity we estimate the ratio of sample width to spacing between voltage probes from images such as shown in Fig. 1. Following the general approach for extended voltage probes we use the center-to-center distance along the current path ($L$) as the sample length and the distance between voltage probes perpendicular to the current path as the sample width ($W$). The sheet resistivity $\rho_{xx}$ is then calculated as  $\rho_{xx}=R_{xx}(W/L)$. The uncertainty in actual current and voltage distribution within our specimens may place an error on the estimated value of $\rho_{xx}$ of less than 30\%.

The resistivity is measured as a function of gate voltage $V_g$ applied between graphene and the degenerately doped silicon substrate. Special care is taken not to collapse the devices electrostatically, as applying gate voltage $V_g$ of either sign leads to an attractive force between the flexible suspended graphene \cite{bunch, electrostatic} and the gate. The observation of graphene collapse at $V_g=20$ V in similar samples leads us to limit the range of applied gate voltages to $\pm 5$ V throughout our experiments. Following Bunch \emph{et al.} \cite{bunch}, we estimate the force acting on our typical device \#1 at $V_g=\pm 5$ V as $F=\frac{\epsilon_0 \epsilon^2 LW V_g^2}{2(d_0+d_1 \epsilon)^2}\sim3\times10^{-8}$~N, where $d_0,d_1=150$ nm are thicknesses of the remaining and etched SiO$_2$ and $L,W\sim3$ $\mu$m are the length and the width of the device. Using simple mechanics, we estimate the maximum strain $\varepsilon$ in graphene to be in the range $V_g=\pm 5$~V as $\varepsilon \sim 0.5(\frac{F}{EtW})^{2/3}\sim5\times10^{-4}$, assuming a Young modulus $E$=1 TPa and a thickness $t=$0.34~nm \cite{bunch}. We deduce that this strain level does not significantly affect electronic transport in graphene.

The blue line of Fig. 2a shows the low temperature resistivity $\rho_{xx}$ of sample \#1, measured as a function of the gate voltage $V_g$. We observe the Dirac peak, indicated by a maximum in the resistivity, at the gate voltage V$_D$  close to zero. The small reproducible fluctuations in $\rho_{xx}(V_g)$ are consistent with universal conductance fluctuation, typically seen in mesoscopic devices \cite{melinda,ucf}.  The carrier density $n$ is determined via Hall effect measurements as $n(V_g)=B/e\rho_{xy}(V_g,B)$, where $B$ is the applied magnetic field. The gate capacitance of the device is calculated as $C_g=n(V_g)e/(V_g-V_D ) \sim 60$ aF$\mu$m$^{-2}$. \cite{novoselov,yuanbo}  The measured capacitance is close to the value  $C_g\sim$ 47$\pm$5 aF$\mu$m$^{-2}$ expected for graphene suspended 150$\pm$20 nm above 150$\pm$20 nm of residual SiO$_2$, as calculated using the serial capacitor model. This provides an independent verification that the device is suspended during the measurements. Finally, using the above carrier density, we determine the electron mobility $\mu=1/ne\rho_{xx} \sim$ 28,000 cm$^2$V$^{-1}$s$^{-1}$ at $n=2\times10^{11}$ cm$^{-2}$.  This is comparable to the best reported values for unsuspended devices at the same density \cite{yuanbo,novoselov,ong,yanwen}. Thus, despite removing the substrate, at this stage the scattering in graphene is not significantly reduced, which leads us to the conclusion that it is caused by residual impurities absorbed on the graphene surface.

Further mobility enhancement requires removal of the remaining impurities. This is accomplished by sending a large current through the device.  For unsuspended samples, this current annealing was demonstrated to heat the graphene sheet locally to an estimated $T\sim600$ C and to desorb most of the residues remaining on the surface of the device from the fabrication steps. While current annealing has been shown to improved the quality of electrical transport in unsuspended devices, the treatment did not lead to significant mobility enhancement \cite{bachtold}.  Most likely, impurities permanently trapped at the interface between graphene and the substrate are responsible for this lack of improvement. Suspended devices, on the other hand, are not be subject to such limitations, since impurities from both sides of the graphene sheet are free to desorb.
Current annealing is implemented by ramping the current across the device up to a predefined setpoint, waiting for several minutes, decreasing the current to zero and remeasuring the electrical transport properties of the specimen. The procedure is applied repeatedly until changes appear in the gate response of the device, which start to occur only at very large current densities of $\sim 2 \times 10^8$ A/cm$^2$, estimated assuming a graphene thickness 0.34 nm.

\begin{figure}
\includegraphics[width=90mm]{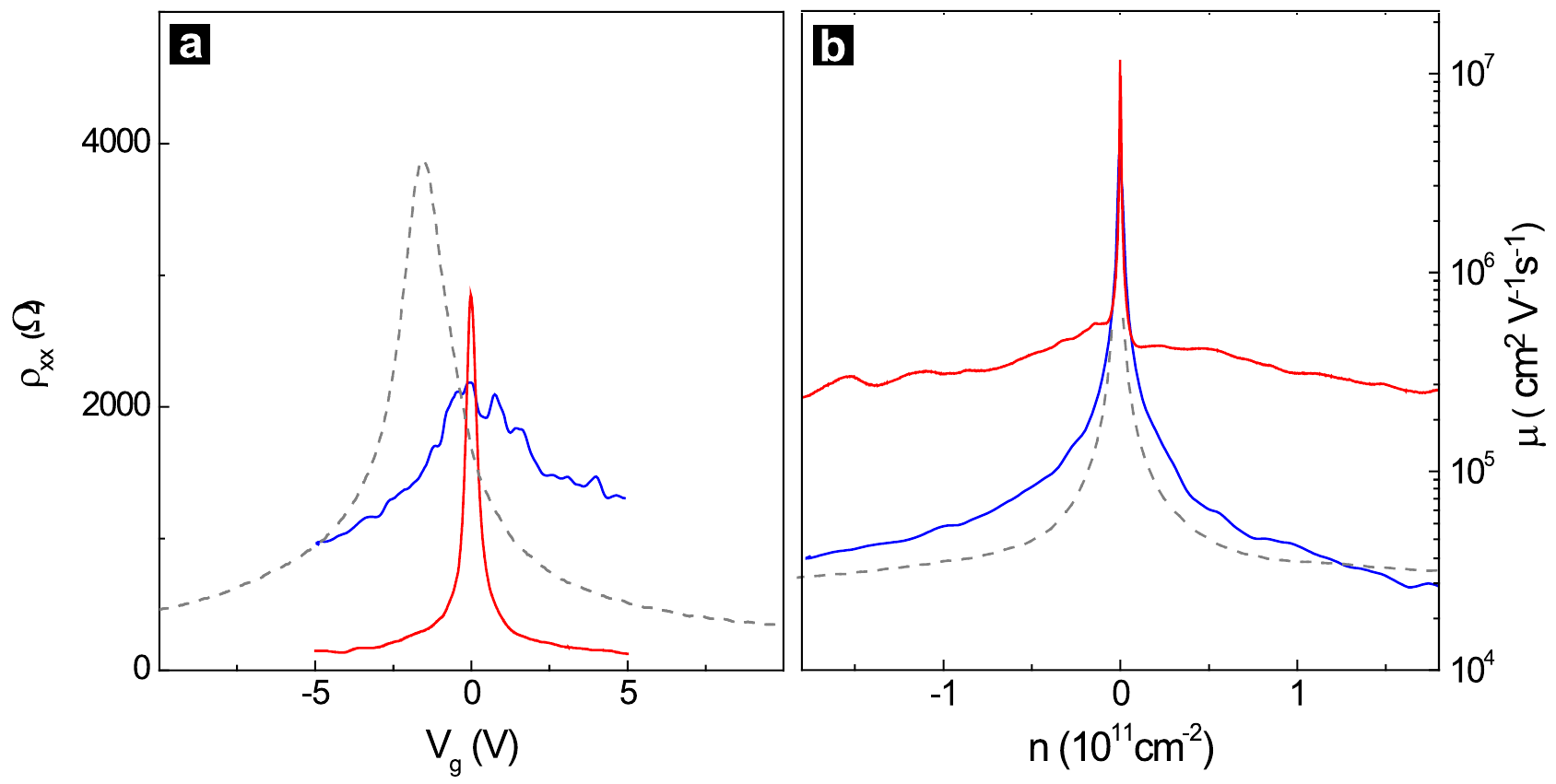}
\caption{(a) Measured four-probe resistivity $\rho_{xx}$ as a function of gate voltage $V_g$ for the device \#1 before (blue) and after (red) current annealing; data from traditional high-mobility device on the substrate (gray dotted line) shown for comparison. The gate voltage is limited to $\pm$5 V range to avoid mechanical collapse. (b) Mobility $\mu=1/en\rho_{xx}$ as a function of carrier density $n$ for the same devices.}
\label{fig2}
\end{figure}

For every device measured, current annealing leads to a remarkable difference in the transport properties compared to the initial state, which we illustrate using device \#1 as an example. Upon current annealing, the resistance of sample \#1 decreases by more than a factor of 8 for voltages away from the Dirac point. At the same time the width of the Dirac peak reduces by about a factor of 20, while the maximum resistivity of the device hardly changes (Fig. 2a). These large changes reflect a greatly improved sample quality. We quantify this improvement via three different measures: carrier mobility, width of the Dirac peak and the onset field of Shubnikov deHaas oscillations.

Our first measure of sample quality is carrier mobility $\mu$ evaluated at high electron density, where $\mu$ saturates. In unsuspended devices, the mobility ranges between 2,000 and 25,000 cm$^2$V$^{-1}$s$^{-1}$ with $\mu\sim25,000$ cm$^2$V$^{-1}$s$^{-1}$ at $n=5\times$10$^{12}$ cm$^{-2}$ being the highest value reported in the literature \cite{ong,yanwen,novoselov}. Due to the gate voltage limitation in our devices we measure the mobility at a smaller density $n=2\times$10$^{11}$ cm$^{-2}$, where the highest reported $\mu$ is about 30,000 cm$^2$V$^{-1}$s$^{-1}$ (Fig. 2b, dotted line). This value is comparable to the mobility of 28,000 cm$^2$V$^{-1}$s$^{-1}$ (Fig. 2b, blue line) in the suspended sample \#1 before current annealing. Upon current annealing, the resistance decrease in sample \#1 translates into an increase of mobility to 230,000 cm$^2$V$^{-1}s^{-1}$ (Fig. 2b, red line) measured at our highest density of $n=2\times$10$^{11}$ cm$^{-2}$. Every suspended device exhibits mobilities higher than 60,000 cm$^2$V$^{-1}$s$^{-1}$ after annealing. Our peak mobility of 230,000 cm$^2$V$^{-1}$s$^{-1}$ represents an improvement of about a factor of 10 over values reported in the literature so far, and is the central result of this work.

In addition to the mobility enhancement, we notice that the Dirac peak of suspended and annealed samples is very narrow compared to both that of suspended devices before annealing and traditional substrate supported devices. We argue that the width of the Dirac peak is related to the charge inhomogeneity inside the sample. As has been demonstrated recently, at small charge densities the graphene breaks into mesoscopic puddles of hole and electrons \cite{yacoby}. The mechanism causing the formation of puddles is debated \cite{dassarma,nomura,geim_intr}, but it is accepted that the presence of puddles changes transport characteristics, resulting in a broadened Dirac peak. We quantify the changes in sample quality by measuring  $\Delta W_{Dirac}$, defined as twice the carrier density at which the resistivity decreases by a factor of two from its maximum value. Such $\Delta W_{Dirac}$ provides an upper bound for the charge inhomogeneity due to puddle formation. In device \#1, for example, the Dirac peak narrows to about 2$\times$10$^{10}$ cm$^{-2}$ (Fig. 2b, red line), an improvement of more than 10 times compared  to the same sample before annealing (Fig. 2b, blue line) and compared to typical high mobility unsuspended devices (Fig. 2b, black dotted line). We remark that the reduced charge inhomogeneity is correlated with enhanced carrier mobility (Fig. 3b). Compared to unsuspended samples (black squares) where a typical charge inhomogeneity is 2-9$\times$10$^{11}$ cm$^{-2}$  while the mobility ranges from 2,000-30,000 cm$^2$V$^{-1}$s$^{-1}$, the suspended and annealed samples (red circles) exhibit both an order of magnitude higher mobility and an order of magnitude lower charge inhomogeneity, following the trend seen in the unsuspended devices.

Finally, we turn to the onset of the Shubnikov deHaas oscillations as a measure of sample quality. In a simple model, these oscillations commence at magnetic field $B_{SdH}$ strong enough for a charge carrier to complete one cyclotron orbit without scattering, which is equivalent to  $\omega_c \tau  \sim 1$, where  $\omega_c$ is the cyclotron frequency and  $\tau$ is the scattering time. In graphene, a semiclassical relation yields  $\omega_c=e v_F B_{SdH}/\hbar(\pi n)^{1/2}$, where $v_F=10^6$ m/s is the Fermi velocity. This results in an estimate $\tau \sim \hbar (\pi n)^{1/2}/ev_FB_{SdH}$. Figure 3a shows the SdH effect in our highest mobility specimen, sample \#1. Oscillations are observed as low as $B_{SdH}\sim 250$ mT (Fig 3a, red line), while no SdH oscillations are observed before current annealing (Fig. 3a, blue line). Other suspended devices exhibit $B_{SdH}$ ranging from 250 to 600 mT, and we estimate $\tau \sim 2\times10^{-13}$ s for the best device at $n=2\times10^{11}$ cm$^{-2}$. On the other hand, in unsuspended devices SdH oscillations at the same density are seen at fields larger than $\sim$ 700 mT, corresponding to  $\tau \sim 7\times 10^{-14}$ s. Therefore, the early onset of Shubnikov deHaas oscillation in the suspended devices is consistent with reduced electron scattering time and thus is indicative of cleaner samples. While the onset of the SdH oscillations is a qualitative measure for sample quality, we cannot deduce directly a quantum scattering time $\tau_q$, since other factors, such as density inhomogeneity, also affect the onset.

\begin{figure}
\includegraphics[width=90mm]{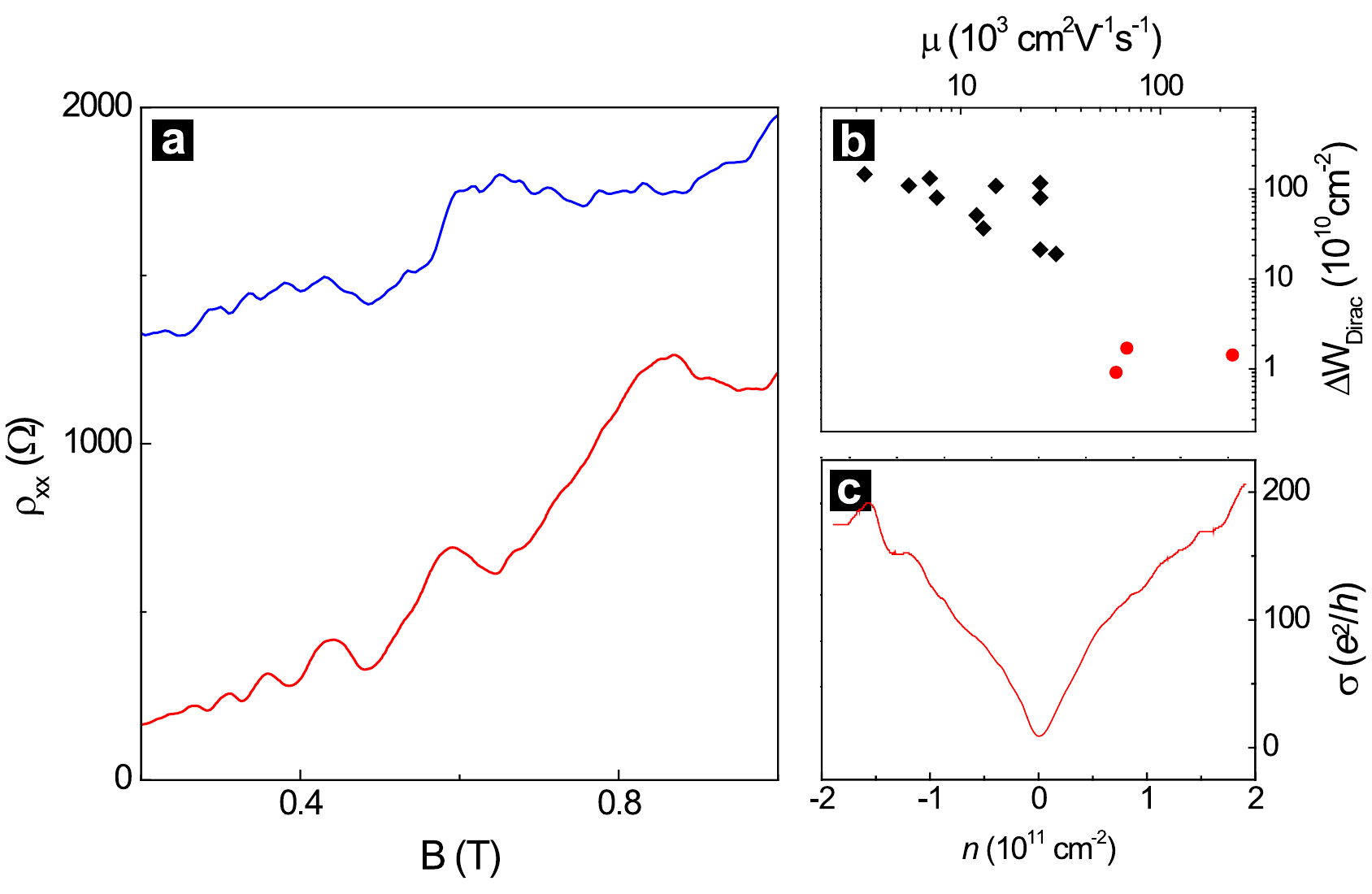}
\caption{(a) $\rho_{xx}$ component of Hall resistance as a function of magnetic field for the suspended sample \#1 before annealing (blue) and after annealing (red) at $n=2\times10^{11}$ cm$^{-2}$ and $T\sim5$ K. (b) Full width at half maximum of the Dirac peak $\Delta W_{Dirac}$ plotted as a function of device mobility $\mu$ for all three measured suspended devices (red circles) and previously studied devices on the substrate (black squares). (c) Conductivity $\sigma$ as a function of carrier density $n$ for the sample \#1 after current annealing.}
\label{fig3}
\end{figure}

Summarizing the results of our transport measurements on in-situ annealed, suspended graphene samples, we observe a considerable improvement in sample quality measured by the enhanced mobility, reduced sample inhomogeneity and increased scattering time. In particular, we observe about an order of magnitude improvement in carrier mobility and sample homogeneity, while the improvement in the onset field of the SdH oscillations is about factor of 3. Overall, we conclude that our fabrication procedure results in very clean samples containing far fewer scatterers compared to the previously studied substrate supported devices. Interestingly, suspended samples prior to current annealing as well as current annealed but unsuspended samples \cite{bachtold} do not exhibit the aforementioned quality improvement. This suggests that impurities trapped between the SiO$_2$ and graphene are limiting the mobility of the current generation of unsuspended graphene devices.

Finally, we consider the nature of the residual scatterers in our devices. Upon current annealing, the carrier mean free path $l$ in our samples approaches the typical dimensions of the device. Indeed, using a semiclassical relation between the mobility and the mean free path \cite{dassarma} $\sigma=en\mu=\frac{2e^2}{h}(k_F l)$, where $k_F=(\pi n)^{1/2}$, we estimate $l\sim1.2~\mu$m for the sample \#1 at $n=2\times10^{11}$ cm$^{-2}$. Therefore, both the edges of the device and the  electrodes may contribute considerably to scattering. This is consistent with the observed strongly sublinear dependence of the conductivity $\sigma(n)=1/\rho_{xx}(n)$ as a function of carrier density $n$ (Fig. 3c). Such behavior was argued to result from the short-range scattering \cite{dassarma,yanwen}, typically associated with point defects or sample edges. Overall, we speculate that extrinstic sources of scattering may still be the limiting factor in the present geometry and that larger area devices may exhibit even higher mobilities.

\section*{Acknowledgements}
We acknowledge fruitful discussions with and experimental help from Erik Henriksen, Jeffrey Kysar, Andrea Young, Barbaros \"Ozyilmaz, and Pablo Jarillo-Herrero. This work is supported by the NSF (No. DMR-03-52738), NSEC grant CHE-0641523, NYSTAR, DOE (No. DE-AIO2-04ER46133 and No. DEFG02-05ER46215), ONR (No. N000150610138), FENA MARCO, W. M. Keck Foundation, and the Microsoft Project Q.



\begin{references}
\bibitem{o2etch}
We use 6 seconds long oxygen plasma etch, with power at 50 W and oxygen pressure 200 mT.
\end{references}
\end{document}